# Integrating Cybersecurity Frameworks into IT Security: A Comprehensive Analysis of Threat Mitigation Strategies and Adaptive Technologies

Amit Lokare, Vanguard[1], USA

Shripad Bankar, Comcast[2], USA

Padmajeet Mhaske JPMC[3], USA

**Abstract**
The cybersecurity threat landscape is constantly actively making it imperative to develop sound frameworks to protect the IT structures. Based on this introduction, this paper aims to discuss the application of cybersecurity frameworks into the IT security with focus placed on the role of such frameworks in addressing the changing nature of cybersecurity threats. It explores widely used models, including the NIST Cybersecurity Framework, Zero Trust Architecture, and the ISO/IEC 27001, and how they apply to industries including finance, healthcare and government. The discussion also singles out such technologies as Artificial Intelligence (AI) and Machine Learning (ML) as the core for real-time threat detection and response mechanisms. As these integration challenges demonstrate, the study provides tangible and proven approaches to tackle framework implementation issues such as legitimate security issues, limited availability of funds and resources, and compliance with legal requirements. By capturing current trends and exposures, the findings promote strong, portfolio-based and risk-appropriate security approaches adjusted for organizational goals and capable to prevent advanced cyber threats.

## 1. Introduction
### 1.1 Cybersecurity in the Modern Digital Landscape
The digital environment has expanded at an unprejudiced rate as advancements in technology reorient the contemporary interpersonal, interorganizational, and international relations. This greatly connected world, as much as it presents a wealth of possibilities for development, has opened the door to some of the most complex cyber threats. With more critical systems being transferred to the online and the cloud being adopted as the standard in modern applications, cybersecurity becomes the final line of defense in protecting digital structures.

As the nature of the cyberspace changes at the present with high speed, so does the forms of cyber threats. Malware, ransomware, phishing and advanced persistent threats (APTs) are the end-point tools used by the hackers where they found vulnerabilities in the new-age connected systems. These attacks are not only those individuals but they can target organizations, businesses or even nations' security. A cybercriminal can be an individual or a group of hackers, or possibly even governments; their goals can be similar to that of a business person, just making money, or have political intentions. This is because with the many threats arising from various sources security solutions have to be ever changing to curb the many threats which are arising frequently.

Thus, the article will reveal the employed cybersecurity frameworks and method, on which an organization's protection methods rely. They will consist of the discussion of the ideas, parts and benefits including The NIST Cybersecurity Framework, ISO 27001 & CISCC. These frameworks provide information on what one should consider susceptible, what protection ought to be secured in place and how the Method of inspection to discern security vulnerabilities and threats, and a Method of protection against them ought to be implemented, can be carried out. In particular, the focus will be made on how these frameworks work to secure IT security structures and what impact they have on risk management.

### 1.2 Evolving Cyber Threats

However, since digital systems constantly develop, the many strategies and techniques used by cyber actors also change. Due to trends in making systems more integrated and closely connected through the use of cloud systems there are more demands for new and complex cyber threats present. From scams and identity theft to hacking personal computers and corporate networks, and state-sponsored cyberterrorism, the number and the types of cyber threats have multiplied. In particular, one of the important things to note in this process is the growth of the complexity and selectivity of threats.

New studies have brought attention to potential risks that are becoming increasingly high for digital identity systems, which are the basis of many contemporary security mechanisms. Digital identities are one of the most important aspects of online transactions, user authentication and management of personal data making them most vulnerable to cyber criminals. Sheik et al. [2] detail the threats on evolving digital identity systems that gives insight into the challenges offered to cyber attackers. Because people now use digital personas in daily activities, opponents use complex tactics including credential stuffing, phishing alongside man in the middle attacks to compromise the systems.

Old school security solutions which were once deemed to be able to protect against basic cyber threats are increasingly proving to be insufficient. New threats are becoming more complicated and therefore require the best security measures that can neutralize these threats in advance. Experts have consequently advocated for a cyber security model that incorporates core security technologies such as the artificial intelligence technology for threat detection coupled with advanced security models that have covered all the general aspects of security models, including the human aspect.

Knowledge of these changing forms of cyber threats reveals the approaches towards building robust protection for businesses operating in the increasingly hazardous digital environments of the modern world.

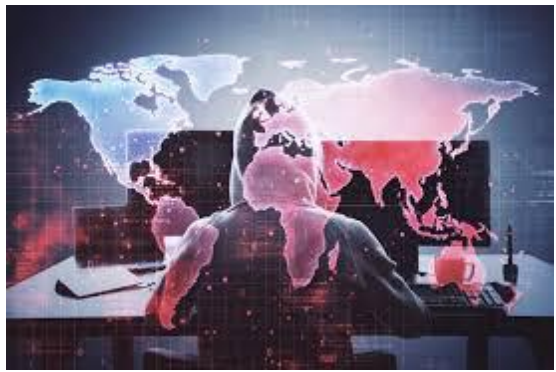

**Figure 1**: *The image of a hacker (Shutterstock, n.d.) highlights modern cyber threats.*

## 2. Understanding Cybersecurity Frameworks
### 2.1 Cybersecurity Framework Basics

Cybersecurity framework therefore refers to a framework which comprise of principles, practices and standards that organization use to protect their information systems, networks and data from cybersecurity threats. These frameworks prevent the creation of cybersecurity approaches that are non-systematic, non-iterative, and can't be successfully replicated – which is precisely what's needed for a rapidly evolving threat landscape in computer crime.

The need for cybersecurity frameworks is anchored in the fact that they help an organization create a level of uniformity and reduce all risks derived from cyber risks. Through instrumentalist, these frameworks provide structures for those organizations to put in place security measures that will be in line with the set standard, minimize risks that might lead to a cyber attack and improve resilience in the case of an attack.

Key reasons for adopting cybersecurity frameworks include:

**Consistency**: Frameworks are designed to make sure all implements of security are similar for all the different systems and processes of the business making it have a coherent security structure.
**Risk Management**: They enable organizations consider the risks and make appropriate decisions on the allocation of funds to avert the highest risks to the business.
**Compliance**: Most of the cybersecurity frameworks reflect legal and regulatory standards, as well as industry standards, thus helping organizations meet cybersecurity standards within their legal environment.
**Continuous Improvement**: Frameworks contain procedures for assessing and reviewing security controls as well as recommending updates and improvements to organizational security systems in response to emerging threats.

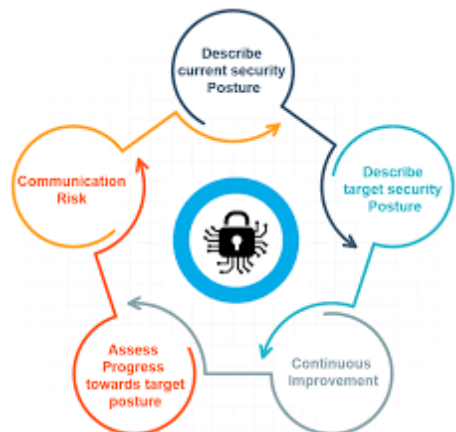

*Figure 2*: *The cybersecurity framework (image retrieved from Edureka, 2018)*

**Frameworks as a Foundation for Consistent and Proactive Cybersecurity**
Frameworks of cybersecurity help applicable organizations have the working tools which enable them to undertake coherent and preventive security measures. These frameworks provide a starting point for organizations to build a strong cybersecurity framework, which will apply universally across their operations Trade-offs for specific activities can then be made consistently with the organizations overall cybersecurity posture.

**Proactive Threat Mitigation**: ZTA has brought up the principles of continuous user and device verification meaning there is no pretrusted identity and that all the identities' access request should be validated upfront [4].

**Data Protection and Monitoring**: Frameworks also highlight the importance of datafication and dataveillance referring to the presence of big data and surveillance methods throughout cybersecurity mentioning live threat monitoring and identification. Risk control is an important part of data governance that makes use of data analytics together with ongoing monitoring in the identification of possible risk openings that further extend to becoming risk threats waiting to be leveraged by the attackers [5].

Also frameworks such as the National Online Informative References (OLIR) program are other approaches that ensure that businesses are provided with consistent forms of information that they can work with to enhance their cybersecurity by providing informative references that organizations follow in regards to cybersecurity and that update them on new and more threatening risks or complex solutions regularly [3]. These sources help cybersecurity professionals to get updated information related to security frameworks that will help them protect their organizations' critical assets from the contemporary threats more effectively.

These cybersecurity frameworks when adopted and implemented, ensure that an organization has a security-first approach to any project, so that risks are constantly assessed, and controls are observed to deter cybersecurity incidents. This is particularly important in the current world where new forms of threats appear constantly, and their complexity increases.

**2.2 Popular Cybersecurity Frameworks**
Given the rapidly changing nature of cyber threats, organizing for security needs to have systematic approaches for addressing problems and prospects. Both of these frameworks are also useful to assist in other areas than merely protecting digital assets, but explaining how to create strong cybersecurity defense. Jointly with that, below is the comprehensive analysis of several of the most popular cybersecurity frameworks in terms of the main concepts and practices.

**1. NIST Cybersecurity Framework (CSF)**

The **NIST Cybersecurity Framework** or CSF is the NIST framework that has proven popular and sought after, which enables organizations to improve in cybersecurity. Originally developed for identifying the critical infrastructure, the NIST CSF being extensible, customizable and compliant with present standards, has been implemented in other sectors as well.

The NIST CSF is built around five core functions:

**Identify**: The first function centers on coming up with an assessment of the organization's cybersecurity exposure. This involves the process of listing those that involve the company's assets, system, data or resources that requires protection. Risk identification process enables the development of a framework within which risks and threats will be managed.
**Protect**: This function defines all the required safeguards for safeguarding the key assets from risks in the future. These are such issues as maintenance of sound access control measures, data encryption, frequent patching as well as security measures against malware and unauthorized access.
**Detect**: It requires constant and active scanning and detection in order to detect that an incident or intrusion is occurring in real-time. In this function, concern is given more on the development of detectors in a view to be able to detect some of these anomalies, intrusions, and any other form of malice.
**Respond**: Controlling the impact of cybersecurity incidents requires having the right strategies for managing incidents. This function addresses how organisations should protect against 'inside' threats, and what to do if an incident is discovered, how the event should be handled and explained; how organisations should limit the consequences of a security breach.
**Recover**: The last function in the NIST CSF relate to recovery activities in order to bring back the functionality that was affected by a cybersecurity threat. This consists of disaster responses and continuity of services through the planning of continuation in case of disruptive incidents, and also the creation of backups in case of service interruption.

The NIST CSF basically provides an effective framework for managing cybersecurity risk at an organizational level with certain degree of flexibility in terms of risk appetite and strategies and goals of the particular organisation that is implementing the risk management process. Its compatibility with other industry standards like ISO/IEC 27001 and regulatory frameworks like GDPR, HIPAA lays it as a golden framework which can be adopted as per the industry's need and size [3].

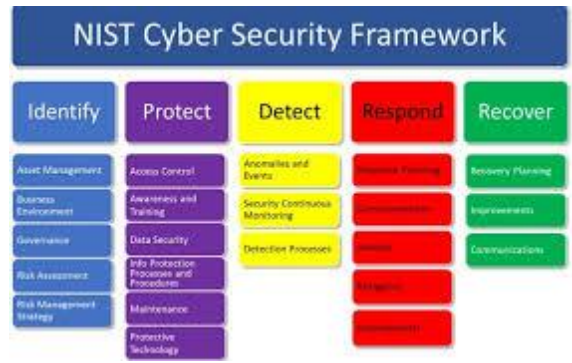

*Figure 3*: The comparative framework (image retrieved from DTS, 2022)

## 2. Zero Trust Architecture (ZTA)

Amidst contemporary cyberspace, Zero Trust Architecture (ZTA) differs as a state-of-the-art concept particularly useful in hybrid computing systems. The Zero Trust model postulates that there cannot be a trusted entity either internal or external to a network. This is a needed shift as perimeter based security models have become more and more ineffective due to more work from home, cloud computing, and advanced hacking.

The core principles of Zero Trust Architecture are:

**Never Trust, Always Verify**: This principle supports constant affirmation of user identities, gadgets, and applications that intends to connect to the organizational resources. There needs to be authentication at every viewpoint, not only at the time of the login. The result is that nobody gets to work with the data without undergoing approval from other parties in the organization.
**Least Privilege Access**: Zero Trust is centered on the idea of least privilege where access is granted to the barest minimum needed for the user and his device. This is important in order to avoid a corrupted entity getting access to information or networks of an organization.
**Micro-Segmentation**: Exemplary of the Zero Trust model, the network is broken down into various compartments, so that an attacker who is able to breach one particular compartment cannot easily move to the next compartment. This segmentation reduces contact and keeps potential foe away.
**Continuous Monitoring and Analytics**: Monitoring of network traffic, users' actions, and system events is continuous because Zero Trust presupposes scanning for risks and threats. Instead of searching for known patterns, behavioral analytics and machine learning are utilized to identify scheduling that may suggest a breach is taking place.

Zero Trust works well when implemented in counteract to insider attacks and the new generation of attacks including spear phishing, lateral movement and data extraction. The primary set of principles of the Zero Trust model is identity and access management (IAM), encryption, and real-time monitoring, which makes it better suited for the modern context where the threats can be more diverse and harder to predict than in traditional, closed environments, such as those promoted by cloud computing and remote work principles [4].

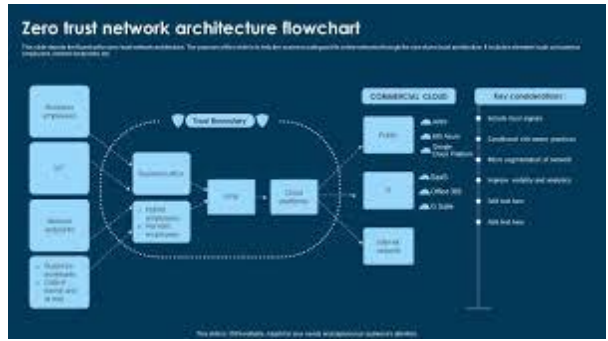

*Figure 4*: *The Zero Trust architecture flowchart (image retrieved from SlideTeam, n.d.)*

## 3. NIST Cybersecurity Framework and Industry Standards

The NIST Cybersecurity Framework also has the advantage of being developed in conjunction with other proven industry standards that provide organisations with guidelines on the direction that they should be taking their security and a road map to ensure compliance with regulatory standards. For example, NIST's framework is frequently paired with:

**ISO/IEC 27001**: This ISO standard for ISMS provides a framework for the governance of organization's information assets and protection from cyber risks. When an organization sign up to ISO 27001 it means that such organization follows best practices in terms of confidentiality, integrity and availability.
**COBIT 5:** The framework used to control IT security and risk is known as the Control Objectives for Information and Related Technologies or COBIT . Measuring up to the NIST CSF therefore holds much value for an organization's IT division, particularly since COBIT is all about getting IT in harmony with organizational goals, creating effective risk management, admitting to and dealing with compliance issues.
**GDPR and HIPAA**: These frameworks offer complements or specific prescriptions on how data should be protected or secured. When an organization implements the NIST CSF, it can be sure that its cybersecurity measures meet legal prerequisites of such privacy regulations as GDPR in EU or HIPAA in the USA.

The European Commission has admitted that the NIST Cybersecurity Framework is quite flexible, which serves well in different regulations. This allows not only to establish a solid cybersecurity program in an organization, but also satisfying compliance and risk control demands put by regulations adopted in various industries, which is important in the healthcare, finance, and critical infrastructure sectors, for instance [6].

Having a good protection system is so vital that cybersecurity frameworks help organizations to design good protection systems. Been the case with NIST Cybersecurity Framework and Zero Trust Architecture being a great place to start and possible to implement to the letter. NIST CSF is more generic and general for most sectors and organisations, it taking a risk based approach whereas Zero Trust offer a rich, layered framework suitable for highly fluid and distributed environments particularly in the contemporary age of cloud and remote working. These frameworks combined with other industry standards help organizations to link cyber security effort with international benchmarks and regulatory compliance thus strengthening the organization's protection from changing and growing threat landscape.

## 2.3 Framework Integration in IT Security

The adoption of cybersecurity frameworks into the IT systems is crucial in creating a comprehensive protective security culture among today's organizations. This integration is important to avoid the situation when security practices are performed separately in isolation, as a result of different departments' work. Implementation of these frameworks also requires the understanding of each of the components in the framework besides the current of the IT environment while applying the best practices to make the integration achievable.

### *2.3.1* How to Integrate Cybersecurity Frameworks into IT Infrastructure

Incorporation of CSFs into IT system can be done by means of several steps starting with the planning and assessment of the IT environment to the implementation of the planned activities and the constant control over their effectiveness. In the following subsections, based on the information provided in the references, there is described the integration process:

**Assessing Current IT Infrastructure**:

However, any cybersecurity framework integration has to begin with an evaluation of current IT systems. This includes learning about the risks to their infrastructure including; key assets and systems and the local regulatory structures. This ultimately helps in choosing the right framework that suits an organization in this case [9]. For instance, a critical infrastructures firm may find it suitable to hire a more detailed and complete framework, NIST CSF.

**Mapping Framework Components to IT Infrastructure**:

Once this is complete, the next step is to align the components of the chosen framework (for instance Identify, Protect, Detect, Respond, Recover of NIST CSF), with the IT structure. This mapping allow each part of the framework to match the organization security requirements. One framework of this style is Depend on Nothing: The Zero Trust Architecture (ZTA) model, where traditional perimeter security is replaced by micro-segmentation along with continuous identity and access management (IAM) across the entire IT structure [4]. Likewise, the controls which covers data protection measures including encryption and segmentation can be executed in accordance with the "Protect" function of the NIST CSF.

**Continuous Monitoring and Adaptation**:

Implementation of cybersecurity frameworks involves structural change and subsequent sensitization to continually changing threats. Frameworks such as NIST CSF and the overall Zero Trust model must have updates in the defined security measures regularly due to newly emerging threats. Information security operation also involves the identification of the detection and response stage where incorporation of SIEM tools could be useful in automating. For instance, while using behavioral analytics to implement the Zero Trust model, any attempts to access the system by an unauthorized person are intercepted and dealt with in real time [9].

**Automation of Security Operations**:

If specific security operations are automated, which include threat detection, access control and incident response, then the interconnectivity of the frameworks is improved significantly. Automating methods are free from human interference, quick, and the security standards can be implemented systematically across all IT mechanisms. Studying the patterns of these threats and identifying their nature to incorporate AI and machine learning in the frameworks such as NIST or Zero Trust can prevent the forming of threats [11].

## 3. Threat Mitigation Strategies in Cybersecurity

The introduction of ICT into all aspects of life in the modern world has attracted vices that come in the form of cyber threats. Hacktivism and cyber-crime with all its forms from viruses to internet identity theft are in themselves complex threats to organizations, state institutions, and citizens. Cybersecurity industry figures revealed that the incidence of cyber attacks has doubled in the last 10 years and malware alone has targeted billions of systems. Thus, over the recent years it emerged that old paradigms of the cybersecurity that is based on ideas, which are firewalls, antivirus programs, and IDSs, are not enough.

Unfortunately, there's a tendency for cyber criminals to innovate and use better methods that can counter anti-virus and security programs and this means that there is need for more effective solution that are capable of identifying and proactively counter attack security threats as well as keep up with the threats in the ever evolving threat landscape. This has led to the quest for a robust solution that addresses this challenge being directed towards the adoption of a new concept that is making huge waves – machine learning (ML) in cybersecurity systems. ML provides a novel approach to replace the rigid traditionally applied methods that rely on strict rules as a security solution by developing systems capable of learning from large amounts of data and identifying threats, besides which it adapted to new threats that had not come across previously. This makes it a more proactive process and leads to far better scalability than what has been possible with more human focused solutions.

This paper presents a review of the existing threats in cybersecurity and its mitigation approaches with emphasis on the application of machine learning towards the boosting of these strategies. We will discuss the particular types of threats which are still prevalent in today's world to understand spikes of interest in the relevant field, the ML methods applied to contain these threats, and the flaws of current models. Last but not the least we will summarize our work, and then briefly outline future directions of cybersecurity research highlighting the future directions promoting the adaptive and intelligent systems for counteracting the adaptive and dynamic character of cyber threats [12].

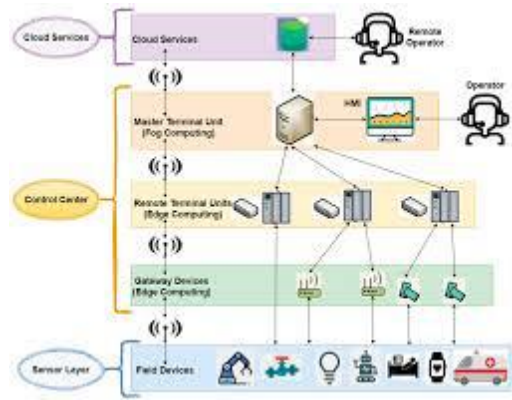

*Figure 5: The unified architecture (image retrieved from MDPI, n.d.)*

### 3.1 Common Types of Cyber Threats

Disasters have changed over the years, complex and difficult to navigate the world wide web this making cyber crimes more unavoidable. The different types of threat in relation to cybersecurity are crucial to provide a basis for threat mitigation. The threats that are out there can in a broad manner be categorized into several classes that include the following; virus, worm, Trojan horse, ransonware, phishing, insider threat among others. However, due to the growth of technologically advanced threat vectors including AI supported attacks as well as zero day threats, there is always new method to the problem.

**Malware, Ransomware and phishing**

Malware is a shortened term for malicious which means it is software created to enter and harm computers or a network of computers. They are viruses worms Trojans and spyware. Ransomware is type of malware that secures a victim's data or records and requests the victim to pay an amount to gain access. Besides, phishing is another traditional threat which aims at getting individuals to share some important information with the intruder by using fake e-mails or other fake web sites.

**Insider Threats**

Insider threats therefore occur when a person with legitimate access to an organization's system or data intends to do harm. These threats can be business motivated, employees with some grudge against the business deciding to embezzle data or can be accidental where the employee becomes a victim of social engineering.

**New and Emerging Threats**

Although cybercriminals are using newer technologies to attack firms, new threats are now appearing. AI integrated attacks means rise of new kind of attack paradigms where adversary have leveraged machine learning to automate and mimic more advance and powerful attacks. Furthermore, attackers can exploit as yet unknown or unaddressed weaknesses — this kind of threats is known as zero-day attacks and it is extremely dangerous for companies.

**3.2 Traditional vs. Modern Threat Mitigation**
Historical development of cybersecurity has brought drastic change from conventional to contemporary threat management. Specific identification or historic techniques including signature mode of detection and the static firewalls primarily work with pattern recognition means. Though such protective systems are good at suppressing set threats, they fail in identifying new threats, for instance, the zero-day threats or threats from artificial intelligence programs as these are evolving and complex threats.

Innovative concepts today employ Artificial Intelligence (AI), Machine Learning (ML), and behavior analysis. Such technologies allow identifying a pattern that deviates from the norm, anticipating probable assaults, and dynamically improving protection against both existing and novel threats. The framework used in this discussion – resilience – also widened the scope for protection to cover not only the prevention of cyberattacks if they were to happen, but also the capacity to quickly bounce back and adapt from such events [14].

In practice, traditional methods remain the basis of defense since modern techniques allow adding comprehensiveness and work flexibility at this level. For instance, basic protection from viruses is offered by regular antivirus programs, whereas the use of artificial intelligence can reveal a slight shift in behavior of the network, which might mean at least a potential threat of breach. Real examples from separate industries such as healthcare and financial are provided to provide deeper understanding of how both approaches complement each other in achieving the best results concerning cybersecurity.

**Table 1: Cybersecurity Threat Detection and Mitigation Techniques**

| Technique | Description | Advantages | Challenges |
|---|---|---|---|
| AI-Powered Threat Detection | Uses machine learning to analyze patterns and detect anomalies | Automated real-time threat detection | High computational cost, potential for false positives |
| Zero Trust Architecture | Verifies every user and device at each access point | Minimizes attack surface, enhanced control | Complex implementation across large networks |
| Deception Technologies | Deploys fake assets to mislead attackers and gather intel | Provides intelligence on attack methods | Resource-intensive, risk of detection by attackers |
| Behavioral Analytics | Monitors user and entity behavior for unusual activities | Detects insider threats, advanced anomalies | Requires large datasets and continuous monitoring |
| Data Encryption | Secures data by encoding it for storage and transmission | Protects sensitive information from breaches | Key management complexities |
| Threat Intelligence Sharing | Collaborative sharing of threat data among organizations | Informed decision-making and proactive defenses | Privacy concerns, trust issues between organizations |
| Incident Response Plans | Predefined protocols for responding to cyber incidents | Reduces downtime, organized recovery efforts | Requires regular updates and testing |
| Cyber Resilience Frameworks | Combines prevention, detection, and recovery strategies | Ensures business continuity and adaptability | Integration with existing systems can be difficult |

[14]

## 4. Case Studies – Real-World Implementations

This makes optimal practical effectiveness of theoretical strategies and frameworks in cybersecurity the true purpose of the theories. This chapter provides examples of how threat mitigation strategies defined earlier are applied in various industries and implemented on the operational level. Such real-life case studies help to explain the practical use of cybersecurity and reveal the issues which may be encountered during its implementation, and show the pros of implementing state-of-art ant cyber threats measures.

From these case studies, this paper seeks to fill the existing literature gap by exploring the practices applied in different organizations in order to shed light on how best organizational practice can be implemented to address certain issues. The topic of this chapter is to provide an overview on how strategic cybersecurity vision translates into planned practical solutions.

### 4.1 Cybersecurity Frameworks in the Financial Sector

It will be pertinent to state that the financial sector remains one of the most attractive sectors within the cyber criminal realm given the nature of information and wealth, that is processed by this sector. Unfortunately, the risk profile of financial institutions has change with the steady evolution of the institutions to fit the digital environment. Since a large number of cybercrimes are evolving in terms of complexity, the importance of robust and flexible cybersecurity frameworks raises significant concerns. This is very much the case, especially with technology companies and especially Financial Technology (fintech) firms that are critical to the financial chain but have relatively less expertise in handling cybersecurity threats.

#### 4.1.1 Evolving Cybersecurity Threats in the Financial Sector

The threat environment in financial industry is dynamically changing, and this process is promoted by the following factors: technological progress, the growth of connecting societies, complication of cyber threats. Writing for CSO, Marican et al. (2022) note that there instances of cyberattacks on technology startups and especially those in the financial technology industry because they do not have well-developed cybersecurity frameworks as is the case with well-established financial institutions. The lack of a broad cybersecurity maturity model for technology startups only worsens the situation in this segment of the financial industry. Sometimes, startup companies may lack the necessary resources and experience to protect themselves from today's complex cyber threats, as many of them are driven by the concept of high growth rates and product innovation [16].

The literature review conducted by Jain et al. (2023) comes to the conclusion that the transition from traditional types of crimes to cybercrimes is a characteristic feature of the present state of affairs in the field of finance cyber-safety. This shift can be linked to financial technology from mobile apps and the now-famous financial tools such as Blockchain and AI services. These generally are new technologies that are creating new and improved ways whereby the hackers can gain access to the systems' vulnerabilities, always at a pace that is much faster than the speed at which even legislation and existing security measures can evolve adequately. This is explained by the fact that for some reason technological change is continuing at a much faster pace than our rules and even laws that, more often than not, are unable to respond to the new threats that flow from such developments as noted by Jain et al [16].

Furthermore, insider threat remains a very influential threat within the financial area. As observed by Marican et al. (2022) insider abuse is a common type of account abuse that can result to high levels of financial and reputational loss. This is because organizations are susceptible to any wrong deeds from any of their personnel who have access to the information or systems of an organization. These are a threat hard to notice and prevent, meaning financial institutions need to enforce a set of internal controls, security procedures and monitoring, and staff training to avoid risks within the financial institution [16].

### *4.1.2* Human Element in Cybersecurity: A Critical Factor

Again, cybersecurity is not just a technology problem but also a people problem. The growing role of human participation in carrying out cyberattacks through either phishing attacks or insider threats means that human behavioral design is an important key to the implementation of organizational cybersecurity initiatives. Cyber security cost of Europe increased since agency,ENISA,said in a 2015 report cybercrime is nearly constantly enabled by social engineering which creates the entry point and makes those targeted or infected by viruses, Trojans and worms active participants [16].

The best way to deal with the human factor in cyberspace is essential training that reveals employees to current threats and the necessity of following security procedures. Lenders have to handle the challenge of cybersecurity education, with cybersecurity becoming everyone's concern excluding the personnel of the IT department.

### *4.1.3* Challenges and Gaps in Existing Cybersecurity Frameworks

One of the identified issues is the lack of the single and the all-encompassing cybersecurity model, which could be implemented to the financial companies, especially the new generation Internet-based technology companies and the fintech organizations. Frameworks like the NIST Cybersecurity Framework are handy; however, they are better designed for mature and enterprise-style banks rather than the innovative and constantly evolving fintech firms.

In the words of Goodwin (2022) there are many financial institutions that have no legally binding requirements for the execution of cybersecurity based on voluntarily adopted frameworks such as the NIST. It implies that, although some organisations can choose to adopt the best practice frameworks, others may not do the same; effectively giving the sector an uneven security position. In particular, differing paradigms combined with insufficient regulation in the sphere of finance increases the problem of protection from the new threats in the sphere of cybersecurity. Much more can be expected from the existing and forming regulatory structures to set and enforce cybersecurity requirements and measures [16].

Not only is there a great deal of regulatory ambiguity, there are also a lot of technological difficulties. The implementation of cloud services has myriad advantages in regard to flexibility and scalability, but where important financial data is stored in public or hybrid cloud environments, there are unique and considerable risks. According to Desai & Hamid (2021), safety measures should be implemented in relation to cloud security; thus, it should meet industry requirements. The issue is how to deal with the risks which are linked to the usage of cloud technologies, and how to take the benefits which are offered by these technologies, at a lower cost and with greater flexibility [16].

### *4.1.4* A Multi-Layered Approach to Cybersecurity in the Financial Sector

Due to the multifaceted character of risks in the contemporary financial sector, several levels are needed to provide protection against a wide variety of threats described in the literature. Such a concept requires the coordination of technical solutions, legal actions, and ethical approaches to protect the institutions, financially.

**Technical Measures**: Credit providers remain under pressure to engage in constant product development and enhancement of their security solutions. This includes enhancing on the advanced security tools which embrace artificial intelligence threat identification, data encryption, and block chain security among others that would go along way in compromising the cyber criminals. In their recent work, Dhingra et al. rightly emphasise the need for developing a range of anti-cybersecurity tools – each capable of addressing different risks [16].

**Legislative Action**: There is no doubt that legislative bodies should assume a predetermining role in the formation of the cybersecurity architecture for the financial services sector. As highlighted by Jain et al. 2023), the evolution of fintech has led to a lacuna in regulatory oversight making it hard to design appropriate cybersecurity strategies. More rigid legislation needs to be enacted for the banks to be pressured into adopting effective cybersecurity and for the adopted measures to uniform the financial industry [16].

**Ethical and Social Responsibility**: Last of all, financial institutions need to weigh in the overall ethical considerations about information security. The few sources of literature available point to CSR practices as being a positive influence towards the provision of transparency, trust and responsibility in the financial services industry.

In the following circumstance: It is only by proceeding to adopt ethical standards in the handling of sensitive data and creating culture of accountability in financial institutions, the financial institutions can enhance their cyber security and apply a more responsible practice in the pursuit of data privacy and security.[16].

### 4.2 Healthcare Industry Cybersecurity

The current world has experienced an expansion in use of intricate technologies and this came with the increase in cybersecurity threats in the health care facilities. These advancements have increased levels of patient care but at the same time made vulnerable patient data and important infrastructure to cyber threats. Some of the effects may include; increased risks to the patient's safety, financial loss and eventual legal action.

#### *4.2.1* Healthcare Data Security Risks

In the current world, hospitals and every other health centers and facilities remain favorite targets of hackers since medical information is so significant. Recorded health information for example is a sensitive information that is most likely to be at risk of identity theft and other financial scams. Another survey that was conducted recently revealed that more than 80% of the healthcare organisations mentioned that they had faced at least one data related incident, out of which 72% were hit by ransomware [17]. Healthcare data is amongst the most sensitive categories of information, and this make attracts attackers. Further, distinct nature and distribution of digital structures substantially across the healthcare systems make them wholly dependent on outdated technologies that are more prone to cyber threats.

#### *4.2.2* Cybersecurity Frameworks in Healthcare

In order to prevent such cybersecurity issues, multiple models were created to help organizations in the sphere of healthcare protect themselves. The NIST Cybersecurity Framework (CSF) is probably one of the most commonly known frameworks for managing and mitigating cybersecurity risks. Components in this framework are for instance the following, the first one being identification of key assets, second being protecting data with vigorous encryption, the third being the capability to detect threats, fourth being the aptitude to respond to incidents and the final one being the ability to recover from disrupts [18]. It is common for healthcare industries to use the NIST CSF as their guidelines because they help when developing policies over cybersecurity.

Other important framework to consider is the HITRUST CSF, which has been developed to meet specifically the needs of the healthcare sector. Thus, HITRUST synchronizes different standards, namely the HIPAA (Health Insurance Portability and Accountability Act) to form a single security architecture.

This framework stresses on the protection of data belonging to patients, protection of Healthcare IT structures and adherence to the Health Information Technology for Economic and Clinical Health Act [17].

The other internationally recognised standard in healthcare cybersecurity is ISO/IEC 27001. It provides procedures for establishing an ISMS to safeguard data in an organization. At the same time, the proposed standard involves comprehensive attention to the issues concerning healthcare IT security, including information networks, disaster communications and patient information security [19].

### *4.2.3* Challenges in Healthcare Cybersecurity

However, these recognized cybersecurity frameworks have not been enough to protect the healthcare system from the major cybersecurity risks. One of the biggest challenges to cybersecurity is the fact that healthcare IT organizations are usually disparate. Most healthcare facilities continue to implement legacy systems and technologies that have minimal or no security features incorporated. The issue of data integration between classic systems and advanced ones is not properly implemented, which leads to programs containing loose ends that can be used by hackers.

Furthermore, many healthcare organisations face inadequate funding for IT security when it comes to healthcare institutions. New research suggests that current funding imperatives do not allow small healthcare systems to invest sufficiently in cybersecurity development. This could result to weak policies that are supported by reduced enforcement of password policies, minimal implementation of encryption and poor event management and response mechanisms [17].

### *4.2.4* Real-World Cybersecurity Breaches in Healthcare

The attacks below are real-life examples of how healthcare organisations have been compromised with their weaknesses exploited. For example, in 2023, the University of California Health System fell victim to a professionally carry out the ransomware-attacks, which affected over 3,000,000 patients' personal health records. Cyber attackers initiated this cyberattack through an email phishing attack then penetrated into the network and encrypted key data while demanding for ransoms from the organizations [17]. Such incidents make it compulsory that hospitals have intricate cybersecurity measures to ensure that their patient's record information is safe from wrong hands and that the institution incurs massive losses by having to pay ransom fees.

The cyber threat attack in the health care sector requires health care organization to embrace defensive and robust cybersecurity strategies. These include frameworks such as NIST CSF, HITRUST CSF, and the ISO/IEC 27001 that healthcare organisations can employ because the number of cyber threats continue to rise. However, a major limitation still exits by means of issues affecting integrated working like disjointed IT infrastructure and inadequate funding. These gaps will have to be filled through cooperation between the regulatory authority and the networks, industry, as well as through developing and improving cybersecurity technologies.

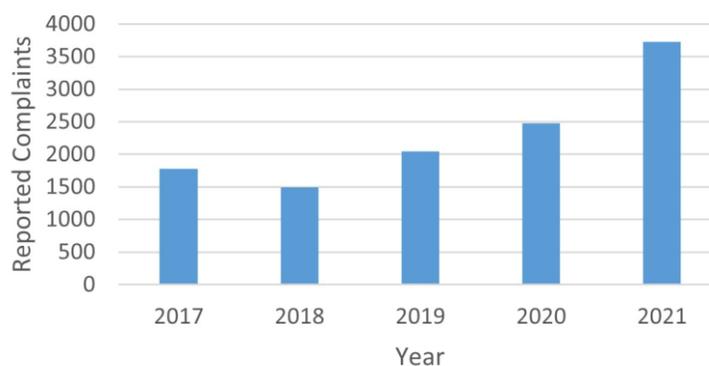

**Figure 6**: IC3 annual report data of ransomware attack (*Image retrieved from NCC Group, n.d.*)

**4.3 Government and Public Sector Cybersecurity**

Government and public sector is considered most vulnerable to cyber attack risks and threats because they process and store highly sensitive information, including national security data, citizen's personal data, and key infrastructure data. Bodies within governments around the world are also at risk from those who are part, or entirely, affiliated with nation-states as well as independent malicious cyber actors targeting weaknesses in the public sector IT environments. Being among the sectors most essential in national strategic and operational management, the protection of government institutions from cyber threats is of utmost importance for the reinforcement of citizens' trust, as well as national and economic security.

*4.3.1* **Cybersecurity Threats to Government and Public Sector Entities**

The reasons for government organizations to be cyber-attack targets are several. These are the information they hold, the sophistication of the IT systems they have and the criticality of affecting their operations. The most common threats include:

**State-Sponsored Cyberattacks**: Nation-states are portrayed to attack governments with objectives of stealing confidential data relating to the nation's security or tampering with the election systems or other critical infrastructure. These assaults are generally well-funded and very sophisticated representing a variety of techniques, some of whch include phishing, unknown duration vulnerabilities and advanced persistent threats (APTs). One of the most recent examples of state-sponsored attack was the SolarWinds attack back in 2020 when the attackers successfully targeted organizations belonging to the US government adopting several of the standard practices used by sophisticated threat actors [21].

**Ransomware Attacks**: Ransomware has become a significant threat for the government organization as the thugs hack the data and encrypt it for which they demand money for the same data to be decrypted. They can lead to a complete shutdown of government functions and incite large-scale devastating economical losses. For instance, the City of New Orleans suffered ransomware attack in 2020 requiring a state of emergency by the city authorities and halting service delivery [20].

**Data Breaches**: Public sector holds a wealth of information about individuals, some of which are filed under taxation information, health information, and social security numbers. Incident to the leakage of this rather delicate information, there are high risks that an individual leaning over such information would engage in acts that would compromise other individuals' identity and that, there would be a general loss of confidence in the relevant institutions. The biggest data breach was seen in 2015 where the federal Government, U.S. Office of Personnel Management lost personal records of at least 21 million people including information on background checks and security clearances, and in the same year [22].

**Insider Threats**: Another form of threat common with governments is known as insider threats. These threats are usually as a result of insiders; employees or contractors who take Part in a company's business but could have malicious intentions. It is true that insider threats can be purposeful whereby the individual, walks away with sensitive information and sells it, or uses it in a wrong way. A familiar case is the Edward Snowden information leak in Jun 2013, a former contractor for the NSA who disclose secret surveillance programs of government [23].

*4.3.2* **Cybersecurity Frameworks for the Government and Public Sector**

Since the government organizations face the risk of cyber threats differently compared to other organizations, developing frameworks more relevant to the public sector is crucial in helping reduce the risks and offer structure. Both of these frameworks aim at improving the security stature of a public sector agency and or organization and demonstrating legal requirements compliance. The most notable frameworks for government cybersecurity include:

**The NIST Cybersecurity Framework (CSF):** This is a government-led best practice framework for managing cybersecurity risks originally prepared by the US National Institute of Standards and Technology (NIST). The NIST CSF is implemented in all the U.S federal agencies and has been referenced by other governments. The framework emphasizes five key functions: Know, Shield, Sensitive, Vessel, and Safeguard, which combined provide the means to address the dynamic risk environment [24].

**The Federal Risk and Authorization Management Program (FedRAMP):** FedRAMP is an initiative of the US government that seeks to define the secure cloud services assessment, authorizations and continuous monitoring processes adopted by federal agencies. FedRAMP confirms that CSPs meet significant security standards that are significant for the security of Govt information and activities in a cloud computing setting [25].

**ISO/IEC 27001**: Although it is not strictly unique to the public sector, the international standard ISO/IEC 27001 for information security management is implemented in a large number of governmental administrations and public organizations in the world. This standard contains direction of implementing an information security management system (ISMS) for an organization [27]. ISO 27001 is essential for public sector organizations that work with personal information to structure data protection from cyber risks.

**The Cybersecurity and Infrastructure Security Agency (CISA) Cybersecurity Framework**: Government services as a critical infrastructure has its own customized cybersecurity framework, which has been established by the CISA an organization under the U.S. Department of Homeland Security. This framework is aimed at exploring and outlining the risks associated with the information system infrastructure of public sector organizations, and the measures that can be taken to minimize the risks of an attack [27].

*4.3.3 Government Regulations and Legislation on Cybersecurity*

To enhance the security of governmental organizations subordinate to the state, different legislative acts have been adopted around the world to control cybersecurity activities in the scope of the public administration. These regulations are there as rules and means of compliance and checking that all the public sector organizations meet the professional standards of data protection and cybersecurity. Key regulations include:

**The General Data Protection Regulation (GDPR):** Effective on the 25th of May, 2018, the GDPR is one of the strictest data protection laws, in force across EU and beyond, affecting not only the private sector but also any public authority that processes data of EU individuals. GDPR requires that respective authorities incorporate sufficient measures to secure personal data, or else face very steep fines [28].

**The Federal Information Security Modernization Act (FISMA):** In the United States, FISMA requires that federal agencies protect their information systems by implementing security standards and other guidance promulgated by NIST. It makes agencies apply cybersecurity controls and conduct risk assessments and high vulnerability discoveries [29].

The Cybersecurity Act of 2015: This was a United States of America law, designed to enhance the security of the nation in the cyber space by promoting sharing of more information between the government and the other entities. It also provides procedure to detect and respond to cyber threats targeting sectors of particular significance such as government departments [30].

## 5. Integration Challenges and Solutions
### 5.1 Integration Challenges with Legacy Systems

The digital environment is constantly changing and growing, and while known, more traditional structures are no longer capable of functioning as a solid foundation of many companies. Made using old technologies and programmers' practices and tools, such systems can contain security holes or architectures that are insufficient for contemporary software systems. This results in a massive difference in security, and thus legacy systems are more prone to cyberattacks compare to emerging counterparts. In this article the author describes the concept of legacy and how security vulnerabilities are endemic to it as well as outlines the potential of modernization to rectify the negative aspects of implementing a legacy

**Overcoming Barriers in Framework Integration**

Historic security models can present interesting problems when integrating into modern cybersecurity frameworks due to several issues that organizations experience during the process. Such challenges are technology, or better said, the lack of compatibility in technologies, resistance to change, lack of resources, and getting through complex legal procedures. New ventures require new frameworks, but the adoption of new frameworks must be accomplished in a strategic manner that scores these issues squarely. The following are principles and tips when it comes to tackling the barriers faced when implementing the framework:

**Technological Incompatibilities**
The issue related to the technological difference between existing legacy systems and the current-generation security frameworks is one of the major challenges faced in the chosen area of research. Current systems are normally developed on old architectures that may not be compatible with new frameworks hence incompatible for integration. Such a scenario translates to poor performance, system halts or in the worst-case scenario being a host to a hacker if not well addressed.

In order to address this issue, organizations have to engage in the acquisition of middleware or integration platforms that connect otherwise incompatible systems. In essence, these platforms can assist in translating previous ones into the supported systems' formats and, therefore, ease integration without a complete overhauling of the whole system [32]. Secondly, the implementation of the integration occurs step by step, it is possible to integrate two or more frameworks in stages to reduce the level of difficulty and the degree of disruption.

**Resistance to Change**
A second factor that make the process difficult is politics within organization that may withstand change, particularly when the structure is well established and accepted by the employees and stakeholders. Common issues to be considered are whether there is the need to disrupt the current organization nets to adopt them, the cost, and the training that may be rendered necessary to promote their use.

Hence, for change to be management facilitated, it is important for it to work on removing resistance and ensuring that both the need for change and the opportunity to change are effectively communicated to the key players and organizational members, that are going to be affected by the change and the method used to affect this is by involving the key stakeholders in the planning and implementation stages. Another way we can facilitate this change is by developing a culture towards cybersecurity and ensuring that everybody undergoes basic training every now and then [33]. Workshops that involve the employees can also create necessary acceptance among them; some of their colleagues can be taken through hand-on demonstrations.

**Resource Limitations**

There is always the problem of time and resource and especially in instances where companies and organizations have to implement the new set cybersecurity frameworks into practice. This statement becomes even more apparent in organizations such as SMEs who quite often have no in-house IT personnel or lack capital for overhauling their systems.

One solution is to look for new outsourced services and cloud-based solutions that can offer good frameworks and security management and can be easily adopted and implemented at low cost [34]. Another possibility is to simply consider integration based on risk, that is prioritizing the integration activities by focusing on the most important risks and starting by addressing those, while slowly extending the integration scope over time. There is always a possibility to turn to external specialists, e.g. cybersecurity consultants, to receive additional help without putting a lot of pressure on the company's internal employees.

**Complex Regulatory Requirements**

Besides, technology and organizational structures impose certain difficulties, not forgetting about the problem in terms of numerous regulations. As already indicated, compliance requirements may be diverse depending on the industry or the geographical location of the organization implementing compliance frameworks that have to meet these requirements.

To mitigate issues concerning the regulatory requirements, the organizations should select the frameworks that will work perfectly for compliance and with compliance aspects of several sectors such as GDPR, HIPAA, and PCI-DSS. Also, meeting with legal and compliance department at an initial stage will also assist in guaranteeing all integration processes fall in line with the standards and laws [32][33]. Corporate compliance status checks will also help in reinforcing the compliance status of the integrated framework periodically to meet the different individual compliance requirements as time goes on.

When it comes to eradication of framework integration barriers, effort, planning and innovation are needed to employ technology, manage change, leverage resources and come to compliance. When these issues are approached tactfully organizations are in a position to harmonize and implement modern security frameworks in their old structures hence becoming all–weather organizations against cyber threats.

**6. Conclusion**

That is why the cybersecurity frameworks are the inescapable and essential instruments in the unceasing campaign in the constantly evolving and intensively hostile cyberspace environment. These fill a gap of giving organizations a framework by which they can protect their IT infrastructure, information and business continuity from threats that are increasing in frequency and complexity. A key area of information protection is the management of risk; NIST Cybersecurity Framework and Zero Trust Architecture as well as ISO/IEC 27001 introduce an organized approach towards the identification of threats and the assessment of organizational defenses, as well as the development of measures to counter threats where identified.

The introduction of other aspects of technology such as the Artificial Intelligent technology (AI) and the Machine Learning have also brought a whole new face when it comes to identification and also the subsequent containing of threats in the organization. These technologies do not only improve the forecast of cybersecurity frameworks but also introduce responses which are excluding intervals that need people, making various processes more efficient. She observed that due to increasing threats, continuous monitoring, more latest technological tools and threat intelligence platforms and behavioural analytics have become essentials for organizations to react and cope new threats which are incontinently emerging.

But the way to the system's full cybersecurity is paved with obstacles. In this case, the legacy system cannot integrate with the current security-related frameworks, hence problems. Lack of resources especially for

SMEs make it difficult for companies to implement suitable high level security measures. Regulatory compliance further increases the problem solving since organizations need to work their way through sector and geographic complexities. Despite these challenges, successful strategies as using middleware technology in system integration, focusing on high risk areas and cross functional cooperation can therefore make way for successful implementation.

Furthermore, the human factor is still very important in the course of any cyber security. Employees our first response mechanism needs consistent training in spotting threats. Awareness of threats to cybersecurity within an organization provides a good defense for those threats in as much as it minimizes human errors infecting corporate systems.

As we move forward, the future of cybersecurity will rely heavily on flexibility, as well as creativity. Frameworks need to mature in order to address segments, for example, the financial one requires efficient real time fraud detection and the healthcare one requires to be secure in storing patient data. This paper argues that regulatory bodies must continue to evolve and expand their understanding of the use of technology to develop coherent and juridical measures. The report highlighted the potential in cross-sector cooperation as threat intelligence as well as cybersecurity practices can be improved by cooperation internationally.

Finally, cybersecurity should be considered a process, and not an implementation since the risk of attacks increases each day. It will be necessary for organizations to continue with future evaluations, refreshing some of the frameworks, as well as funneling cash to advanced technologies so that the defenses against cyberspace threats remain dynamic. In aligning the technical, procedural, and human factors into a means to attain an end, organizations and other stakeholders can be shielded from risks that endanger the growth and sustainability of a digital ecosystem.